\documentclass[11pt]{article}
\def\Journal#1#2#3#4{{#1} {\bf #2}, #3 (#4)}

\def\PLB{{\em Phys. Lett.}  B}

\def\PRD{{\em Phys. Rev.} D}

\def\SNP{\em Sov. J. Nucl. Phys.}
\def\JPL{\em JETP Lett.}
\def\EPA{{\em Eur. Phys. J.} A}
\def\ZPA{{\em Z. Phys.} A}

\def\ko{K^0}
\def\kb{\overline K^0}
\def\bo{B^0}
\def\bb{\overline B^0}

\begin{document}

\begin{flushright}
hep-ph/9910368\\
\end{flushright}

\begin{center}

{\Large\bf
NEUTRAL KAONS AS INSTRUMENT  \\
FOR STUDYING HEAVIER FLAVORS}
\footnote{
To be published in {\it Proceedings of the XIth Rencontres de
Blois, June 27--July 3, 1999, Blois, France.} }

{\large Ya.I. Azimov
\footnote{e-mail: azimov@pa1400.spb.edu}}


{\it Petersburg Nuclear Physics Institute,\\
Gatchina, St.Petersburg, 188350, Russia}

\vspace{0.3cm}

\vspace{0.3cm}
\end{center}

\begin{abstract}

Strangeness oscillations in decays of neutral kaons are suggested
to be used as an analyzer to investigate detailed properties of
heavy flavor hadrons and their decays. Here we briefly explain why,
where, and to what problems this approach may be applied.

\end{abstract}

\section{Introduction}

History of physics provides many examples of how one studied phenomenon
was used for investigating other phenomena. Here we discuss one more
potential case of such a kind. Because of severe space limitations, this
text can be only an extended summary of theoretical work dome in the
direction. For brevity reasons as well, references are given only to
the basic papers published by the present author. A more complete text
may be found in~\cite{azk}. It contains also a longer list of references
including related papers of different authors. Original papers are
strongly recommended for interested readers.

Well-known for many years are strangeness oscillations in time
evolution (and decays) of neutral kaons. They were applied to studies
of the neutral kaons themselves to solve at least three kinds of
experimental problems: 1) measurement of the tiny mass difference
between two neutral kaon eigenstates, unattainable for conventional
mass measurements; 2) identifying $K_L$ as the heavier eigenstate;
3) complete and unambiguous measurements of $CP$-violating parameters
in various neutral kaon decays. As we will explain here, those
strangeness oscillations can (and even should) be applied now to
studies of heavier flavor hadrons (neutral and charged mesons,
or even baryons).

\section{Cascade decays}

Cascade decay is a sequence of decays of both an initial state and
its decay products. It consists of two or more stages which are
usually considered to be quite independent of each other. However,
cascade decays producing neutral kaons may have specific properties.
In particular, time distribution at their secondary stage (i.e., for
the neutral kaon decay) may depend on parameters of the primary decay.

The simplest example is given by decays of charmed particles
(charged mesons or baryons). An essential point here, emphasized
in~\cite{azi}, is that Cabibbo-favored and Cabibbo-suppressed
transitions together produce in such decays a coherent mixture of
$\ko$ and $\kb$.  Then the neutral kaon system evolves and decays.
Its decay-time distribution depends on the initial relative content
of $\ko$ and $\kb$, i.e. on properties of the primary decay.

More complicated are cascades initiated by neutral $D$ or $B$ mesons.
The initial meson here, before its primary decay, develops flavor
oscillations. To the decay moment it transforms into some coherent
mixture of, say, $\bo$ and $\bb$. Its content influences the neutral
kaon content in the system which emerges just after the primary decay.
Thus, the secondary strangeness oscillations coherently continue
oscillations of the initial flavor~\cite{azl}, and we really deal with
coherent double-flavor oscillations. Decay time distributions for the
secondary kaons are sensitive here to both the primary decay and
mixing properties of initial mesons. Moreover, time distributions
of the primary and secondary decays are non-factorisable. Detailed
expressions and their discussion may be found in~\cite{apr,ad} for
$B_d, B_s$ initial mesons and in~\cite{azd} for $D$ mesons.

These examples demonstrate why and how the strangeness oscillations
may be helpful in heavy flavor studies.

\section{Problems for heavy flavors}

In this section we briefly consider some problems in heavy flavor
physics which may be solved by means of secondary kaon oscillations.

\subsection{Identification of meson eigenstates}

Neutral $B$ or $D$ mesons, similar to kaons, evolve in time as
a mixture of two eigenstates. To discriminate them one needs to
use some labels. There are (at least) three possible labels,
just as for kaons: 1) shorter/longer lifetime; 2) lighter/heavier
mass; 3) odd/even $CP$-parity, at least approximate (definition
of the approximate $CP$-parities and discussion of their possible
mode-dependence see in~\cite{apr,ars,azd}~). To achieve complete
identification of the eigenstates one should be able to relate all
these labels with each other.

There seems to be no way for direct experimental connection between
lifetime and mass labels (see discussion in~\cite{azd}~). However,
it is very easy to relate lifetime with $CP$-parity. One should only
compare time distributions in decays of neutral heavy flavor mesons
to final states of definite (and different) $CP$-parities. The
secondary kaon oscillations allow to connect $CP$-parities and masses
of the heavy flavor eigenstates~\cite{apr,ars,azd}, thus completing
their identification. The procedure is similar to how it was  done
for kaons themselves, see discussion in~\cite{azk,azd}. No other
method for complete identification of the heavy flavor
eigenstates has been suggested till now.

\subsection{Parameters of  $~CP$-violation}

$CP$-violation in evolution of neutral $B$ or $D$ mesons and their
decays to final states of definite $CP$-parities can be parametrized
in a manner similar to kaons. However, it was first noticed
in~\cite{ars}, that measurements of the $CP$-violating parameters
can be not complete: they lead to sign ambiguities if one
cannot relate to each other all the eigenstate labels described in the
preceding subsection.  As explained in~\cite{azk}, various sign
ambiguities found later by other authors (see references in~\cite{azk}~)
have the same underlying nature. Thus, oscillations of secondary
neutral kaons, providing complete identification of heavy flavor
eigenstates, at the same tine eliminate ambiguities of
$CP$-violating parameters. Note that measurements of kaon
parameters $\eta$ would also show sign ambiguities if one did
not know that $K_L$ is nearly $CP$-odd and heavier than $K_S$.

$CP$-violation in heavy flavor decays to final states with
neutral kaons has interesting specific features. It combines,
in a coherent way, contributions of two sources: the heavy flavor
itself and the neutral kaons. This is true both for neutral flavored
mesons~\cite{apr,azd,ars} and for any other flavored hadrons
(charged mesons or even baryons)~\cite{azi}. Such properties may
provide new possibilities for studying heavy flavor $CP$-violation
through its "direct" comparison with known kaon violation.

\subsection{Separation of flavor transitions}

A characteristic feature in decays of heavy flavor hadrons, in
difference with strange hadrons, is possibility of several various
underlying quark decays. They generate hadron decay modes with
different flavor changes. Most interesting in the present context
are charmed particle decays of the types $D\to\overline K$
(Cabibbo-favored) and $D\to K$ (doubly Cabibbo-suppressed).
The suppressed decays have the amplitude smallness of order 5\%.
Several modes of decays $D\to K^+$ have, nevertheless, been
observed experimentally.

Decays to charged kaons allow to compare only absolute values of
suppressed and favored amplitudes.  The situation is different
for decays to neutral kaons. Amplitudes of decays to $\kb$ and
$\ko$ appear to be coherent, since the secondary  decays go
to common channels. Therefore, the relative phase of the amplitudes
becomes measurable as well.  It may be measured by means of
secondary kaon oscillations in decays of charmed charged mesons
or baryons~\cite{azi} (detailed formulas for time distributions
see in~\cite{azd} in the limit $\Delta m_D, \Delta\Gamma_D\to0$).
In decays of neutral $D$-mesons the two flavor transitions may
be separated from each other and from initial mixing effects by
means of double-time distributions (over primary and secondary
decay times)~\cite{azd}.

Again, as for the eigenstate identification, such approach is
the only one suggested till now to measure the measurable
relative phase. Detailed knowledge of the suppressed
amplitudes would give new valuable information about CKM-matrix
and $CP$-violation.

\section{Conclusion}

The several examples given  above are sufficient to confirm that the
secondary kaon oscillations may, indeed, have great analyzing
power for heavy flavor physics.

\section*{Acknowledgments}
I thank I. Dunietz, B. Kayser, J.P. Silva and Z.-Z. Xing
for discussions.



\begin{thebibliography}{99}

\bibitem{azk}Ya.I. Azimov, hep-ph/9907260.

\bibitem{azi}Ya.I. Azimov and A.A. Iogansen,
\Journal{\SNP}{33}{205}{1981}.

\bibitem{azl}Ya.I. Azimov,
\Journal{\JPL}{50}{447}{1989}.

\bibitem{apr}Ya.I. Azimov, \Journal{\PRD}{42}{3705}{1990}.

\bibitem{ad}Ya.I. Azimov and I. Dunietz,
\Journal{\PLB}{395}{334}{1997}; hep-ph/9612265.

\bibitem{azd}Ya.I. Azimov, \Journal{\EPA}{4}{21}{1999};
hep-ph/9808386.

\bibitem{ars}Ya.I. Azimov {\it et al}, \Journal{\ZPA}{356}{437}{1997};
hep-ph/9608478.

\end{thebibliography}
\end{document}